\documentclass[english,aps,pre,showpacs,nofootinbib]{revtex4}
\usepackage[T1]{fontenc}
\usepackage[latin9]{inputenc}
\usepackage{amsmath}
\usepackage{amssymb}
\usepackage{esint}
\usepackage{graphicx}
\usepackage{chngcntr}
\usepackage[section]{placeins}
\makeatletter

\@ifundefined{textcolor}{}
{%
 \definecolor{BLACK}{gray}{0}
 \definecolor{WHITE}{gray}{1}
 \definecolor{RED}{rgb}{1,0,0}
 \definecolor{GREEN}{rgb}{0,1,0}
 \definecolor{BLUE}{rgb}{0,0,1}
 \definecolor{CYAN}{cmyk}{1,0,0,0}
 \definecolor{MAGENTA}{cmyk}{0,1,0,0}
 \definecolor{YELLOW}{cmyk}{0,0,1,0}
 }

\usepackage{nicefrac}\usepackage{dsfont}%\usepackage{xfrac}
\usepackage[normalem]{ulem}\@ifundefined{definecolor}
 {\usepackage{color}}{}

\long\def\/*#1*/{}

\makeatother

\usepackage{babel}
\begin{document}

\title{Statistical Properties of Pairwise Distances between Leaves on a Random Yule Tree}

\author{Michael Sheinman$^{1}$, Florian Massip$^{1,2}$ and Peter F. Arndt$^{1}$}

\address{$^{1}$Max Planck Institute for Molecular Genetics, Berlin, Germany\\
$^{2}$INRA, UR1077 Unite Mathematique Informatique et Genome, Jouy-en-Josas, France}

\date{\today}
\begin{abstract} 
A Yule tree is the result of a branching process with constant birth and death rates. Such a process serves as an instructive null model of many empirical systems, for instance, the evolution of species leading to a phylogenetic tree. However,  often in phylogeny the only available information is the pairwise distances between a small fraction of extant species representing the leaves of the tree. In this article we study statistical properties of the pairwise distances in a Yule tree. Using a method based on a recursion, we derive an exact, analytic and compact formula for the expected number of pairs separated by a certain time distance. This number turns out to follow a increasing exponential function. This property of a Yule tree can serve as a simple test for empirical data to be well described by a Yule process. We further use this recursive method to calculate the expected number of the $n$-most closely related pairs of leaves and the number of cherries separated by a certain time distance. To make our results more useful for realistic scenarios, we explicitly take into account that the leaves of a tree may be incompletely sampled and derive a criterion for poorly sampled phylogenies. We show that our result can account for empirical data, using two families of birds species.

\end{abstract}
\maketitle

%\tableofcontents
\section{Introduction}
The speciation process in evolution can be regarded as a branching process.
One of the simplest stochastic models for a branching process is the so called Yule process \cite{yule1924mathematical,karlin1975first}. 
In this model branches are assumed to split with a constant rate and both resulting branches will evolve independently in time. Starting from one branch, a tree will grow, such that the number of leaves on average increases exponentially in time. In a more general version of the Yule tree each branch can also die and get extinct with a constant rate.  

Despite its simplicity, many phenomena in different fields of science have been successfully modeled using the Yule process \cite{newman2005power,novozhilov2006biological}. Particular examples include statistical properties of the number of species in a genus \cite{yule1924mathematical}, the number of members in protein and gene families \cite{yanai2000predictions,reed2004model} and phoneme frequencies in languages \cite{tambovtsev2007phoneme}. In stochastic modelling of biological evolution, the Yule process is often useful as an instructive null hypothesis \cite{raup1985mathematical,aldous2001stochastic,nee1994reconstructed,nee1994extinction}, even when its assumptions are clearly violated.

As an illustrative example of the branching process we present the reconstructed phylogenetic tree of species in the Siilvidae family of birds 
in the left panel of Fig. \ref{SiilvidaeMatrix}. The basis of such a reconstructed tree is pairwise distances between individual species. The color-coded matrix of such distances for the species is shown in the right panel of Fig. \ref{SiilvidaeMatrix}. The statistical properties of such a matrix for a Yule tree is the focus of our article.

Statistical properties of Yule trees have been intensively studied and much is already known. One of the most useful results is the distribution of the number of leaves on a Yule tree \cite{kendall1949stochastic}. This exact analytical result is widely exploited, in particular, for reconstruction of phylogenetic trees and for estimation of rates of speciation and extinction \cite{harvey1994phylogenies,nee1994reconstructed,nee1994extinction}. Other discrete properties have been studied in Refs. \cite{mckenzie2000distributions,steel2001properties,rosenberg2006mean,mulder2011probability} as well as properties of the distribution of branch lengths \cite{steel2010expected,mooers2012branch}.

Often the pairwise distances between all pairs of species in a group of species is the only available information useful for reconstruction of the evolutionary history of the group. For example, in phylogeny reconstruction, one can estimate the pairwise distance in time between two species (twice the time to their last common ancestor) using the molecular clock approach, together with morphological considerations and information about the fossil record \cite{kumar2005molecular}. Motivated by observations of mitochondrial DNA sequences with no recombination, the distribution of pairwise distances has been studied in Ref. \cite{slatkin1991pairwise} for a tree with discrete generations and a given number of leaves. In this study, the authors use a sort of mean-field approach, ignoring fluctuations in the number of leaves during the growth of the tree, to derive an approximate formula for the pairwise distances distribution on a tree.

Here we present a general method to derive the distribution of pairwise distances and other statistical properties on a continuous random Yule tree of a certain height with given birth and death rates. Using our method, we obtain exact, analytic, closed, non-recursive and compact formulas for the pairwise distance distribution, the distribution of distances to the closest neighbour, the distance distribution in so-called cherries, as well as a more general formula for the distribution distance to the $n$-th closest neighbour.
 
Often, in biological context, one does not have an access to data about all existing species
(i.e. leaves of a phylogenetic tree) \cite{mora2011many}.
Instead, species are incompletely sampled, or might have been subject to a recent massive extinction event \cite{pimm1995future}. As long as the extinction of species is random, both scenarios are equivalent on macroevolutionary timescales. In our study, we take the incomplete sampling explicitly into account, which allows us to 
make statements about the fraction of sampled species, using only the available data.

In the next section we will start with a formal definition of the Yule process and then derive the above mentioned distributions of pairwise distances.
% Our results are exact and analytical.
For illustrative purposes we also present numerical simulations perfectly matching our expectations. 
At the end of our article we 
apply our theoretical consideration to empirical data and 
analyze the speciation process in two families of birds
for which data on speciation times and pairwise distances is available.
One advantage of our approach is that we do not need to reconstruct a phylogenetic tree but can solely work with data on pairwise distances.

%===========================================================
\section{A Yule tree with constant branching and extinction rates and incomplete sampling of leaves}
%-----------------------------------------------------------
\subsection{Definition of the Yule Tree}
\begin{figure}[tb]
\centering
\includegraphics[width= 0.5\columnwidth]{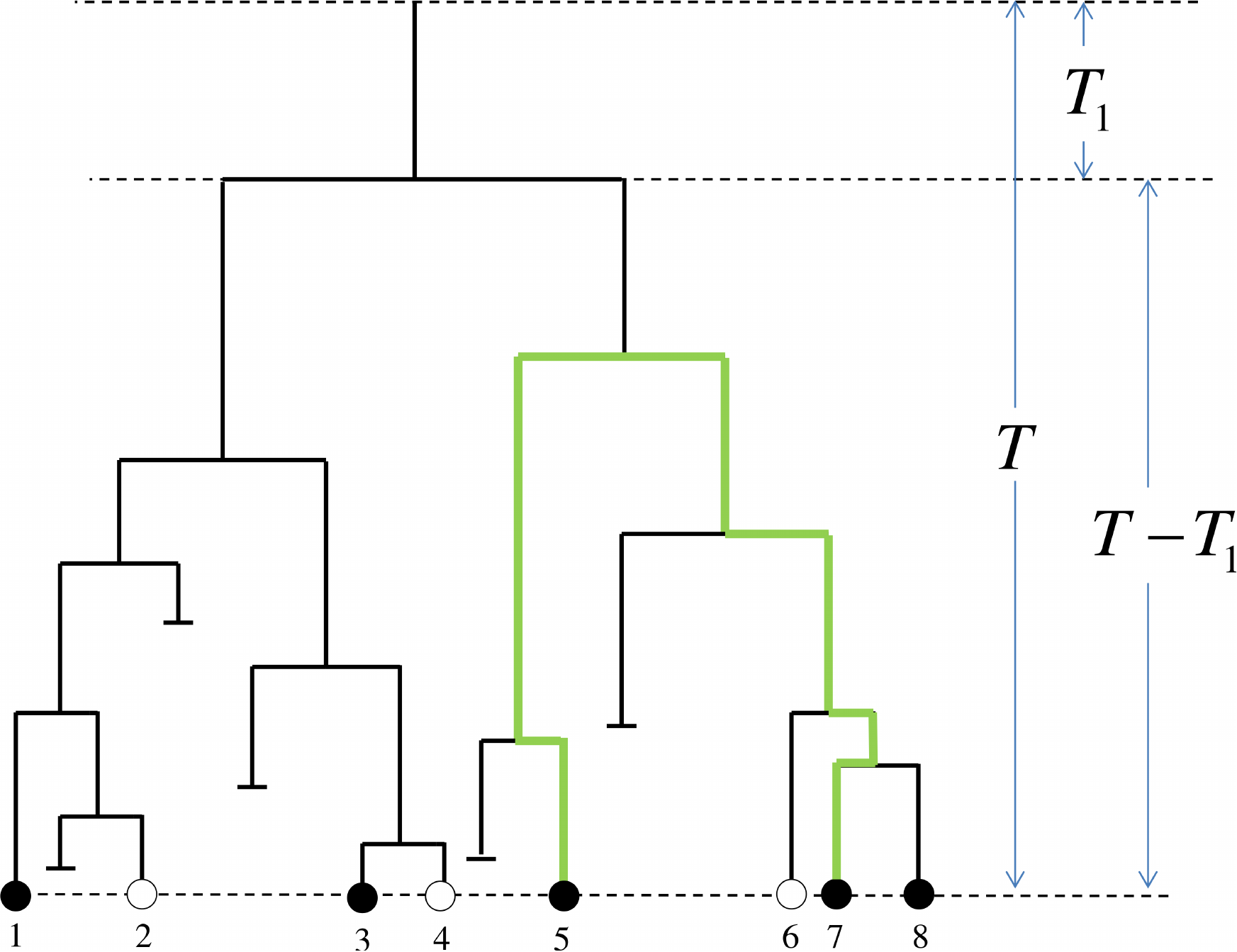}
\caption{%
An example of the rooted Yule tree of age $T$. Filled circles (1,3,5,7 and 8) denote observed leaves. Empty circles (2,4 and 6) denote survived but not observed leaves. Short horizontal lines denotes an extinction event.
After the first branching at time $T_1$ the two resulting subtrees both encompass
$M_1=M_2=4$ leaves. However, the number of observed leaves is 2 (leaves 1 and 3) for the left subtree and 3 (leaves 5, 7 and 8) for the right one. The thick green line denotes the pairwise evolutionary distance between
the two observed leaves 5 and 7. The horizontal dimension is meaningless. In this example for leaf 1 the first closest observed leaf is 3, the second (as well as the third and the fourth) is 5 (or 7 or 8). The tree has two observed cherry pairs: (1,3) and (7,8).}
\label{Tree1}
\end{figure}

A Yule tree is defined as follows \cite{yule1924mathematical,karlin1975first}. At time $t=0$ there is one individual. As time progresses, this individual can branch and give birth to another individual. In an infinitesimally short time interval $[t,t+dt]$, all individuals can give birth to another one, each with the probability $\lambda dt$. The probability of an individual to die in the same time interval is $\mu dt$. We consider an ensemble of trees of age (height) $T$, referring to all existing individuals at this time as \emph{leaves}.
To make the model more realistic, we assume that due to incomplete sampling (or a short massive extinction event) just before the time $T$, each leaf is observed with a certain probability $0 \leq \sigma \leq 1$. The described process is illustrated in Fig. \ref{Tree1}.

\begin{figure}[tb]
\centering
\includegraphics[width= \columnwidth]{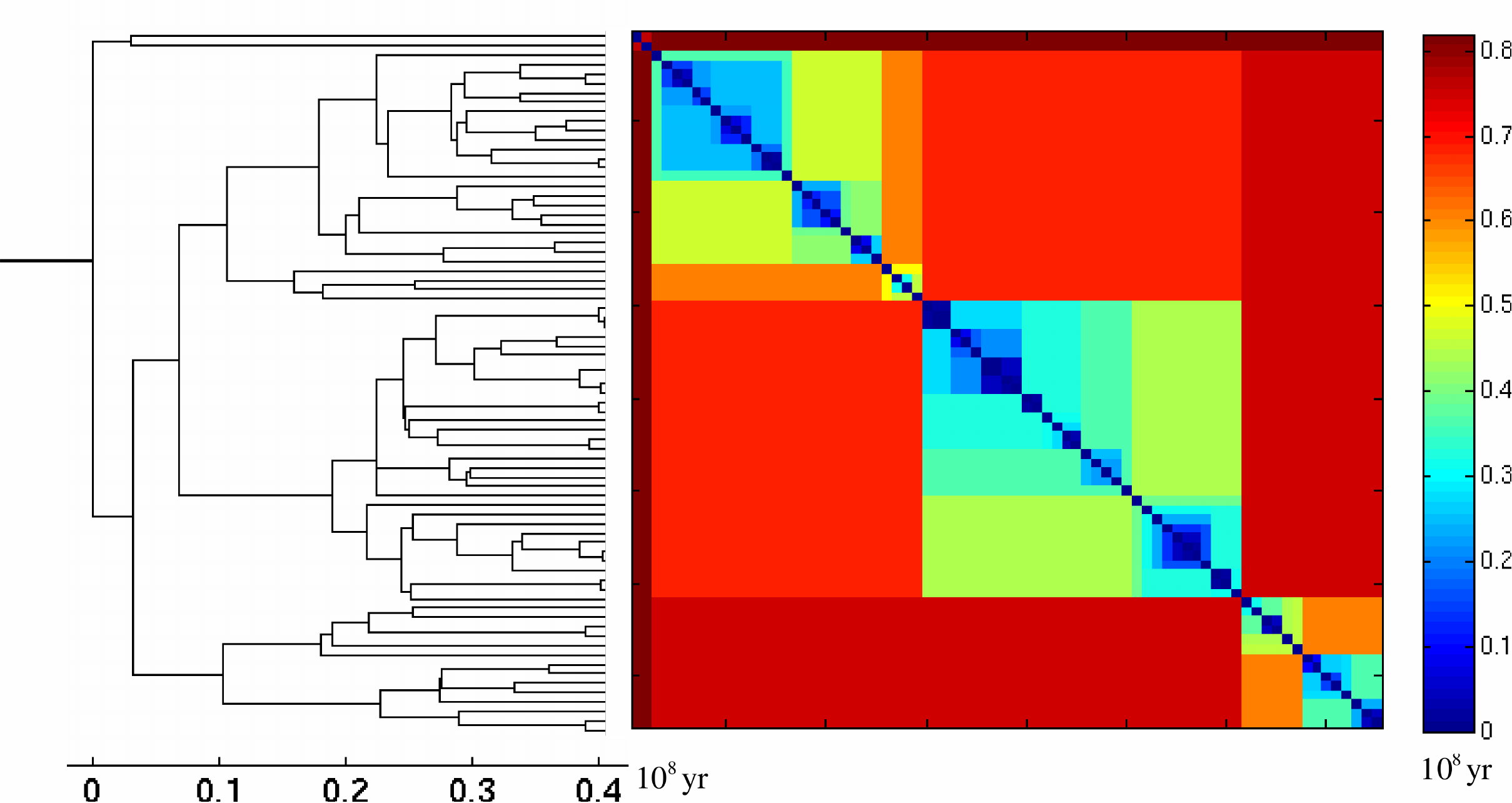}
\caption{%
One of the reconstructed trees for the Siilvidae family of species, taken from \cite{jetz2012global} (left) and its distance matrix (right). The tree includes only the branches which lead to survived and observed leaves.}
\label{SiilvidaeMatrix}
\end{figure}

%-----------------------------------------------------------
\subsection{A Few Useful Results for Random Trees Generated by a Yule Process}

Consider a Yule tree with birth rate $\lambda$ and death rate $\mu$, that have been grown for total time (height) $T$. In the case where all leaves are sampled ($\sigma=1$),
let $P(M|T,\sigma=1)$ be the probability that there are $M$ leaves on a tree of age $T$. Following \cite{kendall1948some},
we can then write the probability that no individual ($M=0$) survives through to time $T$ as
\begin{align}
		%P(0|T,\sigma=1) &=1-\frac{1 -\frac{\mu}{\lambda} }{1 -\frac{\mu}{\lambda}  e^{-(\lambda -\mu )T}}
		P(M=0|T,\sigma=1) &=1-\frac{\lambda -\mu}{\lambda -\mu e^{-(\lambda -\mu )T}}\,.
\end{align}
For $M>0$ we have
\begin{align}
	P(M|T,\sigma=1)&=
	\frac{\lambda -\mu}{\lambda -\mu e^{-(\lambda -\mu )T}}
	%\frac{1 -\frac{\mu}{\lambda} }{1 -\frac{\mu}{\lambda}  e^{-(\lambda -\mu )T}}
	\left[1-\frac{ 1- e^{-(\lambda -\mu )T}}{1 -\frac{\mu}{\lambda}  e^{-(\lambda -\mu )T}}\right] 
	\left[\frac{ 1- e^{-(\lambda -\mu )T}}{1 -\frac{\mu}{\lambda}  e^{-(\lambda -\mu )T}}\right]^{M-1}\,.
\end{align}

We can derive corresponding equations also for the case where species are sampled incompletely. In this case, the probability that 
no species is observed is
\begin{align}
		P(M=0|T) &=P(0|T,\sigma=1)
		+\sum_{m=1}^\infty 	\binom{m}{0} \sigma^0 (1-\sigma)^{m-0}P(m|T,\sigma=1) 
		=\frac{e^{\mu T} (\mu -\lambda +\sigma  \lambda)-e^{\lambda T} \mu \sigma }
			  {e^{\mu T} (\mu -\lambda +\sigma  \lambda)-e^{\lambda T} \lambda  \sigma }
			\label{P0}
\end{align}
and for $M > 0$ 
\begin{align}
	P(M|T)&=\sum_{m=M}^\infty 
	\binom{m}{M} \sigma^M (1-\sigma)^{m-M}P(M|T,\sigma=1)
	=\frac{ \left[e^{T (\mu -\lambda )}-1\right]^{M-1}\lambda ^{M-1}(\lambda -\mu )^2\sigma ^M e^{M T (\lambda -\mu )}}{\left[\lambda  \sigma -\lambda +\mu -\lambda  \sigma e^{T (\lambda -\mu )}\right]^{M+1}}.
	\label{Pn}
\end{align}

Despite these complicated expressions, the average number of observed leaves in a tree of age $T$ is simply given by
\begin{equation}
	\langle M(T) \rangle =\sum_{m=0}^{\infty}m \,P(m|T)=\sigma e^{(\lambda-\mu)T}
\end{equation}
and the average total number of pairs is
\begin{equation}
	\sum_{m=0}^{\infty}\frac{m(m-1)}{2}P(m|T)=\frac{\sigma^2\lambda}{\lambda-\mu}
e^{(\lambda-\mu)T}\left[e^{(\lambda-\mu)T}-1\right].
\end{equation}
The total length of all branches in a Yule tree is given by the integral: 
\begin{equation}
\int_0^T \langle M(T) \rangle dt=
	\int_0^T e^{(\lambda-\mu)t}dt= \frac{1}{\lambda-\mu}\left[e^{(\lambda -\mu )T}-1\right].
		\label{FullLength}
\end{equation}

To derive a corresponding expression for a a tree reconstructed only from incompletely sampled leaves, we note that the
average number of branches at time $t$ with at least one observed descendant at time $T$ is given by
\begin{equation}
	\langle M(t,T) \rangle =e^{(\lambda-\mu)t} \left[1-P(0|T-t,\sigma)\right].
\end{equation}
In the case where $t=T$, we have that $\langle M(T,T) \rangle = \sigma \langle M(T) \rangle$. The average total branch length on the tree of length $T$ excluding the branches which do not lead to an observed leaf is then given by
\begin{equation}
	%\mathcal{T}(T)=
	\int_0^T  \langle M(t,T) \rangle \, dt=
	\frac{ \sigma e^{T (\lambda -\mu )} }{\mu -\lambda  +\sigma \lambda}\ln \frac{\lambda  \sigma +(\lambda -\sigma \lambda-\mu )e^{T (\mu -\lambda )} }{\lambda -\mu }.
	\label{FullObservedLength}
\end{equation}
In the limit of no extinction, $\mu \rightarrow 0$, and exhaustive sampling, $\sigma \rightarrow 1$,  Eq. \eqref{FullObservedLength} is identical to Eq. \eqref{FullLength}. We turn now to calculations of the statistical properties of pairwise distances, using the above formulas.

%-----------------------------------------------------------
\subsection{The Distribution of Pairwise Distances}
\label{Pairwisedistancedistribution}
In a biological context the available data often consist of the pairwise distances separating any pair in a group of species. Commonly these distances are used to reconstruct a phylogenetic tree representing the evolutionary history of a group of species. From such a tree one can then try to estimate rates of speciation and extinction \cite{nee1994reconstructed,nee1994extinction}. 
Here we propose another approach of analysing such data on pairwise distances circumventing the reconstruction of a phylogenetic tree.
 
\begin{figure}[htb]
\centering
  \begin{tabular}{@{}cccc@{}}
  \includegraphics[width=.5\textwidth]{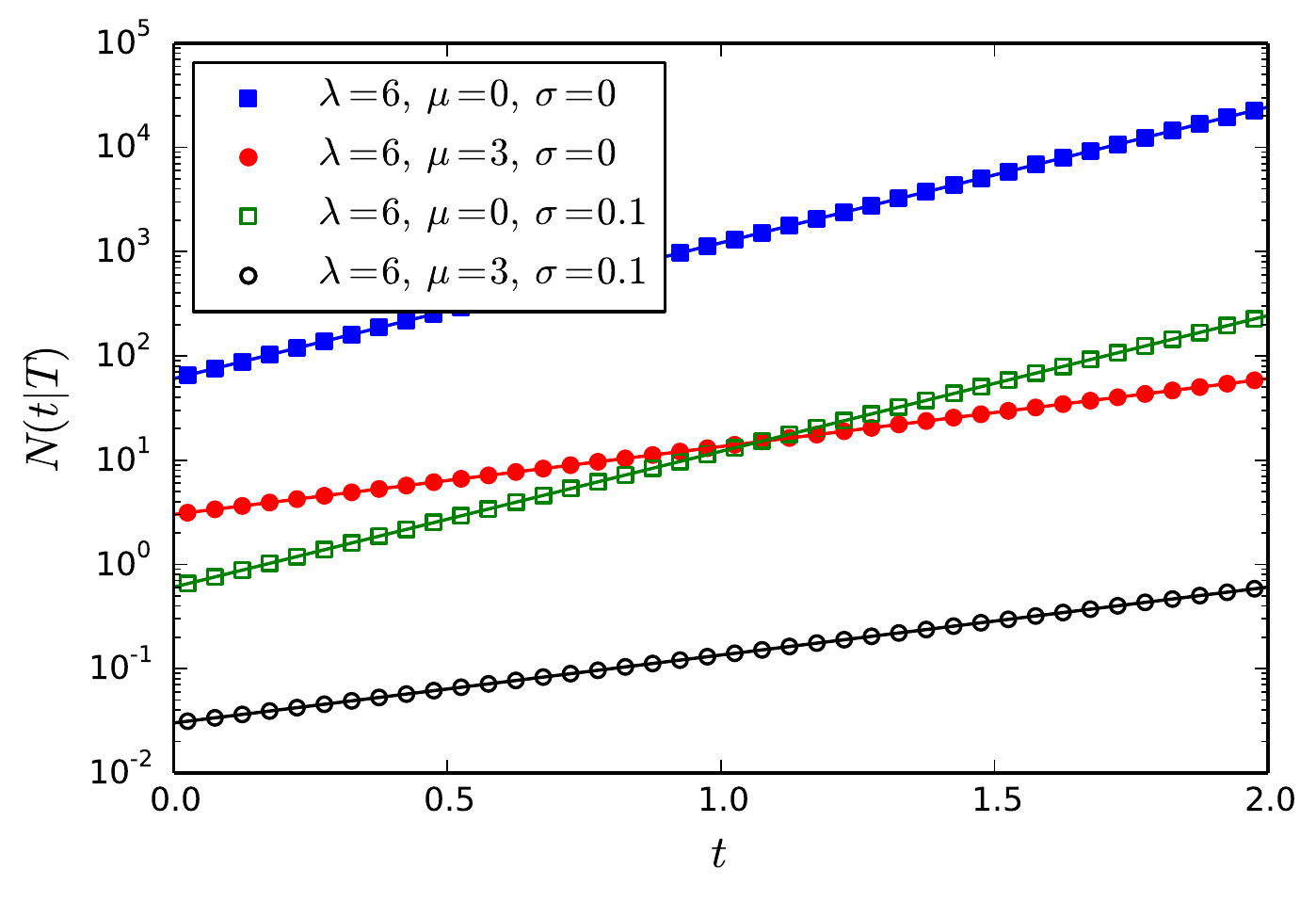}
    \put(-255,160){\bf(a)}&
    \includegraphics[width=.5\textwidth]{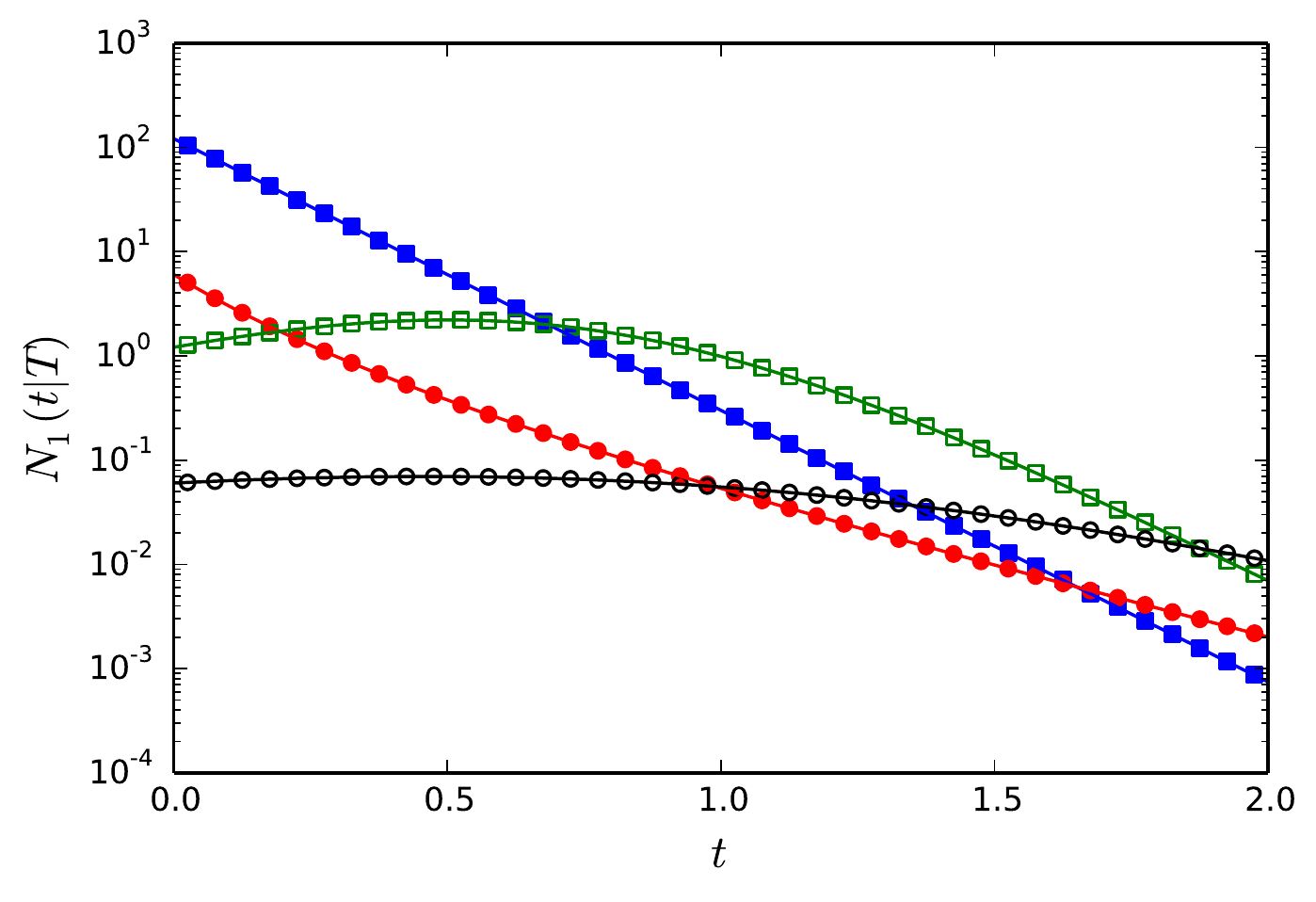} 
        \put(-255,160){\bf(b)}& \\
    \includegraphics[width=.5\textwidth]{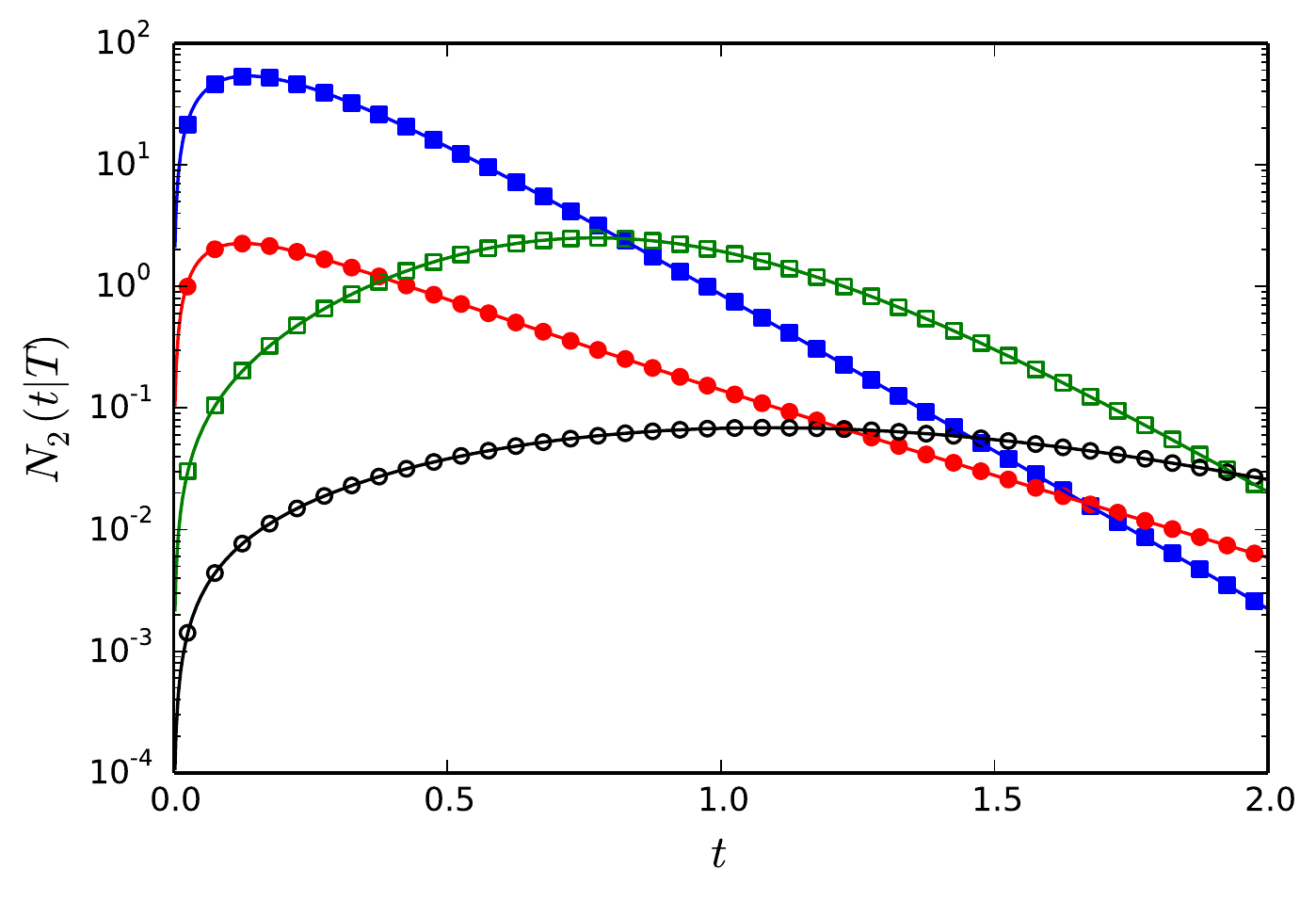} 
    \put(-255,160){\bf(c)}&
    \includegraphics[width=.5\textwidth]{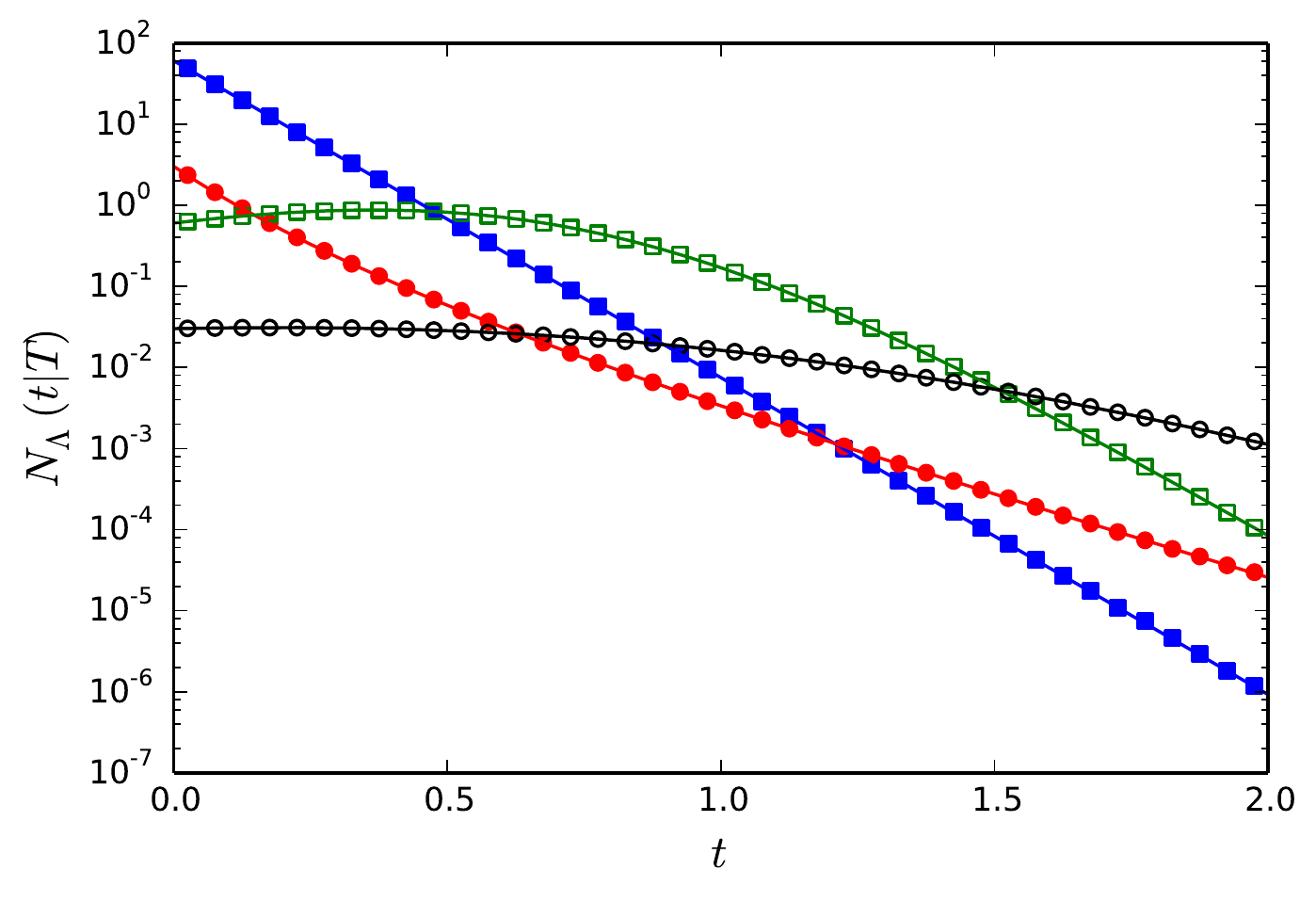} 
    \put(-255,160){\bf(d)}
  \end{tabular}
\caption{Comparison of the analytic results with numerical simulations. Markers indicate numerically obtained data using the following parameters set. $T=1$, $\lambda=6$, $\mu=0$ or $3$ (circles or squares) and $\sigma=1$ or $0.1$ (empty or filled symbols). Lines represent the analytic formulas. (a) Density of number of pairs separated by a certain time, $t$. Lines were obtained using Eq. \eqref{PairwiseDistance}. (b) Density of number of leaves separated by a certain time, $t$ with their closest leaf. Lines were obtained using Eq. \eqref{MinimalDistance} or Eq. \eqref{nMinimalDistance} with $n=1$. (c) Density of number of leaves separated by a certain time, $t$ with their next-closest leaf.  Lines were obtained using Eq. \eqref{NextMinimalDistance} or Eq. \eqref{nMinimalDistance} with $n=2$. (d) Density of number of cherries separated by a certain time, $t$. Lines were obtained using Eq. \eqref{CherriesMinimalDistance}.}
\label{ComparisonToNumerics}
\end{figure}

Let $N(t|T)dt$ be the average number of pairs of leaves on a tree of length (evolution time) $T$, separated by a time distance in the 
interval $[t,t+dt]$, i.e.\,their last common ancestor lived in the time interval  $[T-t/2-dt/2,T-t/2]$. Now consider the branching process as illustrated in Fig. \ref{Tree1}. The first branching happened at time $T_1$ and the two resulting subtrees encompass, say, $M_1$ and $M_2$ leaves, respectively. In this situation one can derive the following recursion relation
\begin{equation}
	N(t|T)=
	\left[2N\left(t|T-T_1\right)
	     +\sigma^2 M_1 M_2 \,\delta\left(t-2\left(T-T_1\right)\right)I\left(0 \leq t \leq 2T\right)
	\right]e^{-\mu T_1}
	\label{PairwiseDistanceRecEquation}
\end{equation} 
where the first part in the summation on the right hand side counts the pairs inside each of the two subtrees
and the second one counts the pairs between them.
The common multiplicative factor, $e^{-\mu T_1}$, expresses the probability that the first branch survives to the time $T_1$ (otherwise, $N(t|T)=0$ ). The function $I$ is the indicator function, defined by:
\begin{equation}
	I( \mbox{condition})=\begin{cases} 
	1 & \mbox{if condition holds} \\
	0 & \mbox{otherwise}
\end{cases}
\end{equation} 
and $\delta(x)$ is the Dirac delta function.
Averaging over $M_1$, $M_2$ (using Eqs. (\ref{P0},\ref{Pn}) with time $T-T_1$) and then $T_1$, which follows an exponential distribution with mean $1/\lambda$, one obtains:
\begin{equation}
	N(t|T)=2\lambda \int_0^\infty N\left(t|T-T_1\right)e^{-\left(\lambda-\mu\right) T_1}dT_1+\frac{\sigma^2 \lambda }{2}e^{\lambda t} e^{-\left(\lambda+\mu\right) \left(T-t/2 \right)}I\left(0 \leq t\leq2T\right).
\end{equation}
In Laplace space one gets:
\begin{equation}
	N(t|S)=2\lambda  \frac{N\left(t|S\right)}{S+\lambda+\mu}+
	\frac{\sigma^2 \lambda}{2}\frac{e^{\lambda t -St/2}}{S+\lambda+\mu},
\end{equation}
where $S$ is the Laplace conjugate variable of $T$.
Solving and inverting the Laplace transform one finally gets the solution:
\begin{equation}
	N(t|T)=\frac{\sigma^2 \lambda}{2} \, e^{\left(\lambda-\mu\right)T} \, e^{\left(\lambda-\mu\right)t/2}  
	\label{PairwiseDistance}
\end{equation}
for $0 \leq t \leq 2T$ and zero otherwise.
Fascinatingly, this distribution is a simple exponential function in $t$. The distribution is cut off at $t=2T$ because in 
a tree of age $T$ two leaves cannot be separated by a time larger than $2T$.
In Fig. \ref{ComparisonToNumerics}(a) 
we show this distribution of pairwise distances for several parameter values together with results of numerical simulations, which match perfectly our 
theoretical expectations.

One can also derive the same result using the following simple arguments. Pairs, separated by a time in the interval $[t,t+dt]$, branched at the time interval $[T-t/2-dt/2,T-t/2]$. The average number of branches in this interval is given by $\lambda e^{(\lambda-\mu)(T-t/2)} dt/2$. The average number of observed pairs from a branch at this time is given by $(\sigma e^{(\lambda-\mu)t/2})^2$. Multiplying the two factors one gets Eq.~\eqref{PairwiseDistance}. However, for other quantities, derived below, the recursive equation approach is more effective.

\subsection{The Distribution of the Minimal-Distance to Other Leaves}
Using the recursive method from Section \ref{Pairwisedistancedistribution} one can also compute other interesting quantities.
For instances in certain situations, the distance separating a leaf to its most closely relative may be estimated more precisely than its distance to other leaves in the tree.
Thus, we might be interested in $N_1(t|T)dt$---the average number of leaves on the tree of age $T$, separated by the time distance between $t$ and $t+dt$ from their most closely related leaf. Interestingly, calculating this quantity lets us make certain statements on the value of the sampling rate $\sigma$. 

To calculate this distribution, we can again write a recursion relation, assuming that the first branching occurred at time $T_1$. In this case one gets the distribution of the minimal distance time in the form
\begin{equation}
	N_1(t|T)=\{2N_1\left(t|T-T_1\right)+
	2 P(1|T-T_1) \left[1-P(0|T-T_1)\right] \delta\left(t-2\left(T-T_1\right)\right)I\left(0 \leq t \leq 2T\right)\}e^{-\mu T_1},
	\label{N1eq}
\end{equation} 
where $P(M|T)$ is the probability to observe $M$ leaves after time $T$, as computed in Eqs. (\ref{P0}) and (\ref{Pn}). In contrast to the recursion relation for the distribution of all pairwise distances, we count a branching point only if $M_1=1$ and $M_2>0$ or $M_1>0$ and $M_2=1$, as expressed by the product $2 P(1|T-T_1) \left[1-P(0|T-T_1)\right]$ in Eq. \eqref{N1eq}.

Averaging Eq. \eqref{N1eq} over $T_1$, one gets:
\begin{align}
	N_1(t|T)=2\lambda \int_0^\infty N_1\left(t|T-T_1\right)e^{-(\lambda+\mu) T_1}dT_1+
	\frac{e^{- (\lambda +\mu )T+ \left({3 \lambda }/{2}+\mu \right)t} \lambda  (\lambda -\mu )^3 \sigma ^2}
	{\left[e^{\frac{ \lambda t}{2}} \lambda  \sigma-e^{\frac{ \mu t}{2}} (\mu -\lambda  +\sigma \lambda) \right]^3}I\left(0 \leq t\leq2T\right).
\end{align} 
The solution of this equation is given by
\begin{equation}
	N_1(t|T)=\frac{e^{\frac{ \lambda t }{2}+ \lambda T+ \mu t- \mu T} \lambda  (\lambda -\mu )^3 \sigma ^2}
	{\left[e^{\frac{ \lambda t}{2}} \lambda  \sigma-e^{\frac{ \mu t}{2}} (\mu -\lambda  +\sigma \lambda) \right]^3}
	%I\left(0 \leq t\leq2T\right)
	\label{MinimalDistance}
\end{equation}
for $0 \leq t\leq2T$ and 0 otherwise.
Results of numerical simulations perfectly match our theoretical expectations (see Fig. \ref{ComparisonToNumerics}(b)).
Interestingly, the function $N_1(t|T)$ from Eq. \eqref{MinimalDistance} possesses a maximum only if 
\begin{gather}
    \sigma<\frac{1}{3}\left(1-\frac{\mu}{\lambda}\right)\leq\frac13
\end{gather}  
and the position of the maximum 
\begin{gather}
	t_{\rm max} \equiv \frac{2}{\lambda -\mu }\ln\frac{\lambda (1-\sigma )-\mu }{2 \lambda  \sigma }
\end{gather}
is in the range $[0,2T]$.
This result is useful for a quick estimation of the data completeness. In particular, a maximum in the distribution of the minimal distance imply that the sampling of the considered tree is not complete and $\sigma<1/3$.

By similar arguments we can also derive expressions for the distributions of second minimal distances, $N_2(t|T)$ (see Appendix \ref{App_2Minimal}) and of the $n$-th minimal distance $N_n(t|T)$ (see Appendix \ref{App_nMinimal}) to other leaves.
The latter quantity is computed to be
\begin{equation}
N_n(t|T)=\frac{ n (1+n) (\mu -\lambda )^3 \sigma  (\lambda  \sigma )^n}{2}\frac{ \left[e^{\frac{1}{2} t (\mu-\lambda )}-1\right]^{n-1} e^{\frac{n t \lambda }{2}+T \lambda +t \mu -T \mu } }{\left[e^{\frac{t \mu }{2}} (\mu -\lambda +\sigma  \lambda )-e^{\frac{t \lambda }{2}} \lambda  \sigma \right]^{n+2}}
%I\left(0 \leq t \leq 2T\right)
\label{nMinimalDistance}
\end{equation} 
for $0 \leq t\leq2T$ and 0 otherwise. 
In Appendix \ref{App_Cherry} we also calculate the distribution of distances in "cherries", i.e. in pairs of  leaves that are adjacent to each other (see Fig. \ref{Tree1} for illustration of cherries): 
\begin{equation}
	N_{\Lambda}(t|T)=\frac{ \lambda  (\lambda -\mu )^4 \sigma ^2}{2 }\frac{e^{\frac{ t \lambda }{2}+T \lambda +\frac{3 t \mu }{2}-T \mu } }{\left[e^{\frac{t \mu }{2}  } (\mu -\lambda  +\sigma  \lambda  )-e^{\frac{t \lambda }{2}  } \lambda  \sigma \right]^4}
	%I\left(0 \leq t\leq2T\right)
	\label{CherriesMinimalDistance}
\end{equation}
for $0 \leq t\leq2T$ and 0 otherwise. 
For illustration purposes we show the distributions for the second minimal distance in Fig. \ref{ComparisonToNumerics}(c) and, for cherries, in Fig~\ref{ComparisonToNumerics}(d).

%------ here -----

\section{Beyond the Averages}
Above results are average expectations. For instance, in Section \ref{Pairwisedistancedistribution} we derive $N(t|T)$, defined as the \emph{average} density number of pairs, separated by a certain time distance $t$, on a tree of length $T$. The average is over many realizations, say $S$ many, of the Yule trees with a given set of parameters $\lambda$, $\mu$, $\sigma$ and $T$. Namely, 
\begin{equation}
	N(t|T)=\left< N^s(t|T)\right>_s=\lim_{S \to \infty} \frac1S {\sum_{s=1}^{S} N^s(t|T)},
\end{equation}
where $N^s(t|T)$ is the density number of pairs separated by a time distance in the interval $[t,t+dt]$ in an individual sample tree number $s$. In reality one often possesses information only about one specific tree $s=1$, i.e. $N^1(t|T)$. Therefore, we are interested not only in the derived averages of $N(t|T)$, $N_n(t|T)$, $N_{\Lambda}(t|T)$ etc. but also their distributions in finite time intervals. The last becomes especially important in the maximum likelihood fitting and model testing. In the discussion below we refer to the distribution of the number of pairs separated by a certain time, $N^1(t|T)$. However, the same arguments can be applied to other quantities, like the $n$-th minimal distance or the distance  in cherries, which we mention above.

Consider an infinitesimal (in practice very small) interval, $[t,t+dt]$, such that $N(t|T)dt \ll 1$. The number of pairs $N^1(t|T)dt$ in this interval is distributed with the mean $N(t|T)dt$. However, in the considered small bin limit, the mean does not represent well the typical value because the distribution of $N^1(t|T)dt$ is not well peaked but possesses a very small probability of having any positive value, while probability of having zero is almost one (see  Appendix \ref{AppNs} and Fig. \ref{NsDistFig}).
\begin{figure}[tb]
\centering
\includegraphics[width= 0.5\columnwidth]{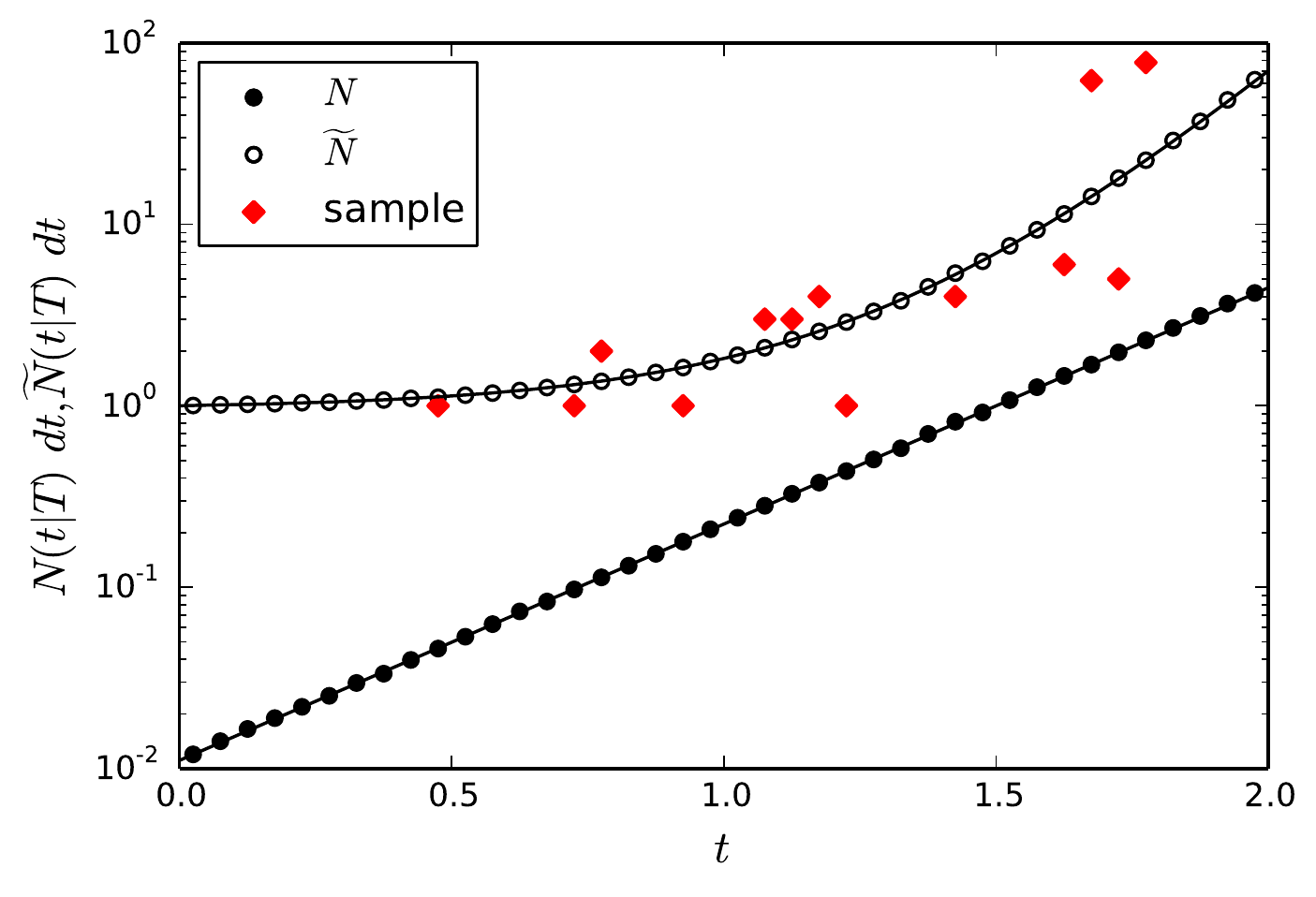}
\caption{The benefit to use $\widetilde{N}(t|T)$ instead of $N(t|T)$ to estimate the parameters of the evolution process in a case of a small dataset. In this plot $T=1$, $\lambda=11$, $\mu=5$, $\sigma=0.01$ and $dt=0.005$. After average over many samples ($S \sim 10^6$ in this particular case) empirical averages of both $N(t|T)$ (full circles) and $\widetilde{N}(t|T)$ (open circles) converge nicely to the analytic formulas. 
The last are given in Eqs. \eqref{PairwiseDistance} and \eqref{Ns}, respectively, and are denoted by the lines in the figure (see the legend). 
However, for a single random tree, $S=1$, the values of $N^1(t|T)$ (diamonds) are highly dispersed (most intervals show zero counts and do not show up in the semilogarithmic plot), such that their fit to the analytic formula of $N(t|T)$ is not expected to lead to a good estimation of the model's parameters. In contrast, the values of $N^1(t|T)$, ignoring the bins where $N^1(t|T)=0$, are well distributed around $\widetilde{N}(t|T)$, although in this example the tree possesses only $19$ observed leaves, such that the data is very poor (only 171 pairs in total).}
\label{NtildaFig}
\end{figure}
Pairs separated by the time in the interval $[t,t+dt]$ branched at the time interval $[T-t/2-dt/2,T-t/2]$. The probability to have a branch in this interval is given by $\lambda e^{(\lambda-\mu)(T-t/2)} dt/2$. Given that there is a branching point in this interval it can lead to different number of leaves. The probability that no observed pairs survive from this branching is given by
$
	1-\left[1-P(0|t/2)\right]^2,
$
where $P(M|T)$ is the probability to observe $M$ leaves on a tree of age $T$ and is given in Eqs. (\ref{P0},\ref{Pn}).
Therefore, the probability that there are no observed pairs separated by the time in the interval $[t,t+dt]$ is given by
\begin{equation}
	\Pr\left (N^1(t|T) dt=0\right)=
	1-\lambda e^{(\lambda-\mu)(T-t/2)} dt/2 \{1-\left[1-P(0|t/2)\right]^2\}.
	\label{PrNs0}
\end{equation}  

In sum, in the small bin limit it is convenient to break the full distribution in two distributions: One comprising only the peak at zero and a second representing all samples with $N^1(t|T) dt \neq 0$. The total average can be broken as follow:
\begin{equation}
	N(t|T) dt=0 \times \Pr\left (N^1(t|T) dt=0\right) + \widetilde{N}(t|T) dt \times \left[ 1-\Pr\left (N^1(t|T) dt=0\right)\right].
\end{equation}

Here $\widetilde{N}(t|T)$ is the average of $N^1(t|T)$ over the tree realizations with $N^1(t|T)>0$. It can be computed to be:
\begin{align}
	\widetilde{N}(t|T)&=\lim_{S \to \infty} \frac{\sum_{s=1}^{S} N^s(t|T)}{\widetilde{S}(t)} = \frac{N(t|T)}{1-\Pr\left (N^1(t|T) dt
	=0\right)} 
	=	\frac{1 }{ dt}
	\left(1+\sigma \lambda \frac{  e^{\frac{ \lambda-\mu }{2}t}-1}{  \lambda -\mu  }\right)^2,
	\label{Ns}
\end{align} 
where $\widetilde{S}(t)={\sum_{s=1}^{S} \left[1-\delta_{N^s(t|T),0}\right]}$ is the number of samples with $N^1(t|T)>0$.
Since, $1-\Pr\left (N^1(t|T) dt=0\right) \ll 1 $, the value of $N(t|T)dt$ is not representative of the expected empirical average of $N^1(t|T) dt$ for finite $S$ and, in particular, $S=1$.
However, the value of $\widetilde{N}(t|T)$, derived above (see Eq. \eqref{Ns}), is representative of the expected empirical average of positive values of $N^s(t|T) dt$. We illustrate this in Fig. \ref{NtildaFig}

\section{Comparison of the derived results to empirical data}
\label{Comparison}
\begin{figure}[tb]
\centering
\includegraphics[width= 1\columnwidth]{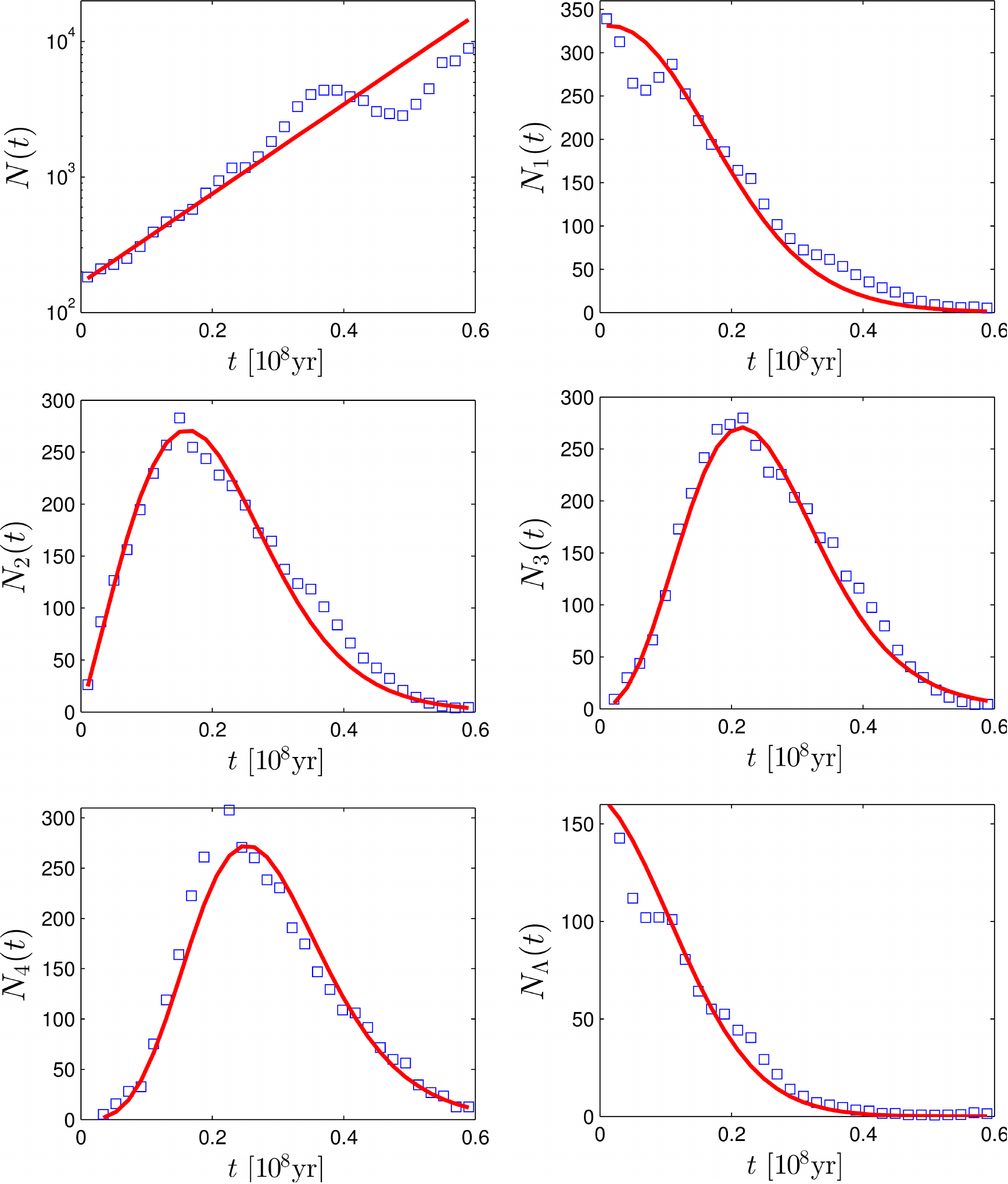}
\caption{Comparison of analytic predictions to the pairwise distances data of Sylviidae family with $M=75$ species taken from the database \cite{jetz2012global} with $t \leq 0.6 \times 10^8\rm{Myr}$. The markers represent the empirical data, while the lines represent the analytic formulas with fitted parameters. (a) Pairwise distance distribution. (b) Minimal distance distribution.(c-e) $n$-minimal distance distribution.  (d) Cherries distance distribution.
The lines are based on following set of parameters: $\lambda -\mu = 15.2  \times 10^{-8} \rm{yr}^{-1}$  and $\lambda \sigma = 4.6 \times 10^{-8} \rm {yr}^{-1}$. For $\mu=0,0.2,0.4,0.6,0.8 \times \lambda$ this corresponds respectively to $\sigma=0.3,0.24,0.18,0.12,0.06$.}
\label{S12} 
\end{figure}

\begin{figure}[tb]
\centering
\includegraphics[width= 1\columnwidth]{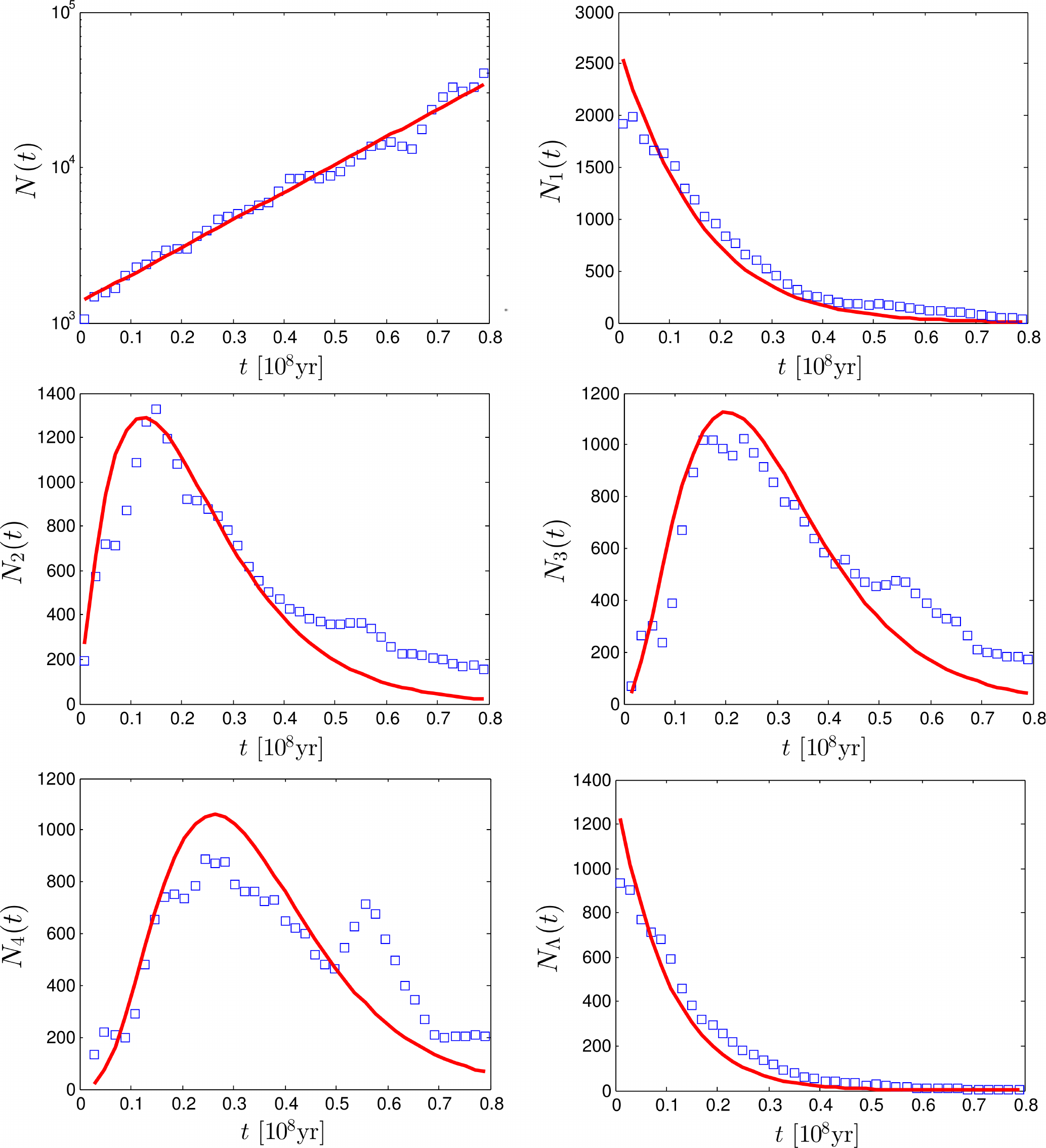}
\caption{Comparison of analytic predictions to the pairwise distances data of Tyrannidae family with $M=460$ species taken from the database \cite{jetz2012global} with $t \leq 0.8 \times 10^8\rm{Myr}$. The markers represent the empirical data, while the lines represent the analytic formulas with fitted parameters. (a) Pairwise distance distribution. (b) Minimal distance distribution.(c-e) $n$-minimal distance distribution.  (d) Cherries distance distribution.
The fit is performed for all points in the figure with $t \leq 0.5$. 
The lines are based on following set of parameters: $\lambda -\mu = 8  \times 10^{-8} \rm{yr}^{-1}$  and $\lambda \sigma = 6.4 \times 10^{-8} \rm {yr}^{-1}$. For $\mu=0,0.2,0.4,0.6,0.8 \times \lambda$ this corresponds respectively to $\sigma=0.8,0.64,0.48,0.32,0.16$.
}
\label{TitTyranRest} 
\end{figure}

In this Section we demonstrate the relevance of the obtained analytic formulas to empirical data, studying the pairwise distances between species in families of the evolutionary tree. For comparison with the derived results we choose $N(t|T)$, $N_n(t|T)$ with $n=1,2,3,4$ and $N_{\Lambda}(t|T)$. The results are presented in Fig. \ref{S12} for the Siilvidae family of birds (see one of the reconstructed trees for this family and its distance matrix in Fig. \ref{SiilvidaeMatrix}) and for the Tyrannidae family of birds in Fig. \ref{TitTyranRest}. For every family we analyze Bayesian sampling of $1000$ trees downloaded from the database \cite{jetz2012global}. Namely, we collect pairwise distances, $n$-minimal distances and distances between cherries of all 1000 trees and plot the histograms of these distances (with the $y$-axis divided by 1000) in Figs. \ref{S12} and \ref{TitTyranRest}.
We fit all the points in a figure using the iterative reweighted least squares algorithm \cite{holland1977robust} in Matlab. Unfortunately, the explicit dependencies on $\lambda$ and $\mu$ in Eqs. (\ref{PairwiseDistance},\ref{nMinimalDistance},\ref{CherriesMinimalDistance}) are insufficient to estimate all parameters. Instead one can estimate from the fit only the effective growth rate, $\lambda-\mu$ and $\lambda \sigma$.  The value of $\sigma$ can be obtained assuming a certain ratio $\mu/\lambda$. In the captions of Figs. \ref{S12} and \ref{TitTyranRest} we present the obtained estimates for $\sigma$ for different assumptions about the ratio $\mu/\lambda$.

Over all, the fits to empirical data look satisfactory and result in a reasonable set of parameters, which roughly agree with the ones given in \cite{jetz2012global}.
This indicates that certain statistical properties of speciation can be well captured by a simple Yule process.
However, in some cases, deviations can be observed. For example, for the Sylviidae family the pairwise distances distribution deviates from the prediction for $t>30$ Myr, while for the Tyrannidae family we observe a clear deviation for distances around $55$ Myr in all our estimates. This  possibly indicate a massive radiation event in the considered family of birds around $27.5$ Myr ago, as already reported in \cite{jetz2012global}.

Interestingly, we can state that the Sylviidae family of birds is currently not well sampled as independently on the assumed ratio of the death rate to the birth rate all estimated values of the sampling fraction $\sigma$ are below 30\%.

\section{Summary and concluding remarks}
In this paper we present a novel method to calculate statistical properties of Yule trees. The method is based on a recursive equations which can be solved using the Laplace transform. We demonstrate the strength of our method deriving formulas for (\emph{i}) average number of pairs separated by a certain time (Eq. \eqref{PairwiseDistance}), (\emph{ii}) the number of most closely related pairs separated by a certain time (Eq. \eqref{MinimalDistance}), (\emph{iii}) the number of  next-most closely related pairs separated by a certain time (Eq. \eqref{NextMinimalDistance}), (\emph{iv}) the number of $n$-most closely related pairs separated by a certain time (Eq. \eqref{nMinimalDistance}) and (\emph{v}) the number of cherries separated by a certain time (Eq. \eqref{CherriesMinimalDistance}).

Our results can be compared to the empirical data using only the information about pairwise distances between leaves of a considered tree. The reconstruction of the tree structure is not required. This is a particular strength of our method because the reconstruction of such trees for a large number of leaves is sometimes problematic. 
In such cases one often considered a posterior distribution of trees which is generated by Bayesian sampling \cite{bouckaert2014beast,ronquist2003mrbayes}. Such a distribution of trees can still be easily analyzed using our method.

We demonstrate the relevance of our results to statistical properties of pairwise evolutionary time distances between biological species. We find that in some cases the speciation process is well described by the Yule model. Significant deviations from the derived distributions are expected to be indicative for massive extinction or radiation events. In the case where the assumptions of the Yule process are justified, we expect our results to be useful for estimation of the incompleteness of the data sampling, i.e. the fraction of observed leaves out of all existing leaves, $\sigma$. However, similarly to the method developed in Ref. \cite{nee1994extinction}, all the derived results depend only on three parameters: $\lambda-\mu$, $\lambda \sigma$ and $\sigma e^{(\lambda-\mu)T}$. Therefore, even knowing those \emph{three} parameters one cannot estimate the values of the \emph{four} unknown parameters: the rates $\lambda$, $\mu$, the height of the tree, $T$ and the sampling fraction, $\sigma$, without an additional assumption about one of these parameters, for instance the fraction $\mu/\lambda$. If it is known that the sampling is perfect, $\sigma=1$, one can estimate both the birth and the death rate. However, in contrast to Ref. \cite{nee1994extinction}, the method presented here does not require the reconstruction of the tree, but is solely based on statistical properties of pairwise distances between the leaves of the tree.

In the general case, one can get an upper limit for the sampling fraction and a lower limit for the birth rate by setting $\mu/\lambda=0$. 
These bounds are expected to be useful for analysis of exponentially growing trees. Such trees can appear in phylogeny when analyzing the evolution of taxa, but also in population genetics when considering an exponentially growing sub-population under the influence of a positive selection.

\acknowledgments
The authors thank M. Mariadassou, P.W. Messer, and M. Vingron for helpful discussions.

%================================================================
\appendix
\counterwithin{figure}{section}
\section{Second-minimal-distance distribution}
\label{App_2Minimal}
Let $N_2(t|T)dt$ be the average number of leaves on the tree of length $T$, separated by the time distance $t$ from their second-most closely related leaf. Then, if the first branching occurs at time $T_1$ and the two resulting subtrees possess $M_1$ and $M_2$ leaves, respectively, one gets the distribution of the minimal distance time in a form
\begin{align}
	N_2(t|T)=&2N_2\left(t|T-T_1\right)e^{-\mu T_1}\nonumber\\
	&+2 \left[2P(2|t/2) \left(1-P(0|t/2)\right) + P(1|t/2) \left(1-P(0|t/2)-P(1|t/2)\right)\right]\nonumber\\
	&\times \delta\left(t-2\left(T-T_1\right)\right)I\left(0 \leq t\leq2T\right)e^{-\mu T_1}.
	\label{NextMinimalDistanceEquation}
\end{align} 

After average over $T_1$ and solving the resulting equation one obtains
\begin{equation}
N_2(t|T)=\frac{3\lambda ^2 (\lambda -\mu )^3 \sigma ^3 \left(e^{\frac{t \lambda }{2}}-e^{\frac{t \mu }{2}}\right)}{\left[e^{\frac{t \mu }{2}} (\mu -\lambda  +\sigma  \lambda )-e^{\frac{t \lambda }{2}} \lambda  \sigma \right]^4}e^{\frac{t \lambda }{2}+T \lambda +t \mu -T \mu }
%I\left(0 \leq t\leq2T\right)
\label{NextMinimalDistance}
\end{equation} 
for $0 \leq t\leq2T$. 
Similarly, one can obtain any third-minimal distance distribution fourth- etc. The general formula for the $n$-minimal-distance distribution is calculated in the following.

\section{$n$-minimal-distance distribution}
\label{App_nMinimal}
Let $N_n(t|T)dt$ be the average number of leaves on the tree of length $T$, separated by the time distance $t$ from their $n$-most closely related leaf. This notation means that $1$-most closely related leaf is the closest one, $2$-most closely related leaf is the second-most closest one etc. Then, if the first branching happens at time $T_1$ and the two resulting subtrees possess $M_1$ and $M_2$ leaves, respectively, one gets the distribution of the minimal distance time in a form
\begin{align}
	N_n(t|T)=&2N_n\left(t|T-T_1\right)e^{-\mu T_1} \nonumber \\
	&+2\left[n P(n|t/2) P_>(0|t/2)+(n-1) P(n-1|t/2) P_>(1|t/2)+...+ P(1|t/2) P_>(n-1|t/2)\right] \nonumber \\
	&\times \delta\left(t-2\left(T-T_1\right)\right)I\left(0 \leq t \leq 2T\right)e^{-\mu T_1}\nonumber \\
	=&\left[2N_n\left(t|T-T_1\right)
	+2\delta\left(t-2\left(T-T_1\right)\right)I\left(0 \leq t \leq 2T\right)\sum_{k=1}^n k P(k|t/2) P_>(n-k|t/2) 
	\right]e^{-\mu T_1}
\end{align} 
Here
\begin{equation}
 P_>(k|T) =\frac{\sigma ^{k+1} (\mu -\lambda )\lambda ^k\left[e^{T (\mu -\lambda )}-1\right] ^k \left(e^{T \lambda } \lambda -e^{T \mu } \mu \right)^k e^{T \lambda }}{\left[e^{T \mu } (\mu -\lambda  +\sigma  \lambda  )-e^{T \lambda} \lambda  \sigma \right]^{k+1}\left[\lambda -e^{T (\mu -\lambda )} \mu \right]^k}
\end{equation}
is the probability to observe more than $k$ leaves on a tree of age $T$ and $P(n|T)$ is given in Eqs. (\ref{P0},\ref{Pn})
After average over $T_1$ and solving the resulting equation one obtains
\begin{equation}
N_n(t|T)=\frac{ n (1+n) (\mu -\lambda )^3 \sigma  (\lambda  \sigma )^n}{2}\frac{ \left[e^{\frac{1}{2} t (\mu-\lambda )}-1\right]^{n-1} e^{\frac{n t \lambda }{2}+T \lambda +t \mu -T \mu } }{\left[e^{\frac{t \mu }{2}} (\mu -\lambda +\sigma  \lambda )-e^{\frac{t \lambda }{2}} \lambda  \sigma \right]^{n+2}}
%I\left(0 \leq t \leq 2T\right)
\end{equation} 
for $0 \leq t\leq2T$ and 0 otherwise, resulting in Eq. \eqref{nMinimalDistance}.

\section{Cherries-distance distribution}
\label{App_Cherry}
%\begin{figure}[tb]
%\centering
%\includegraphics[width= 0.25\columnwidth]{Cherry.pdf}
%\caption{An example of cherry (dotted lines) on a tree.}
%\label{Cherry}
%\end{figure}
A cherry is a pair of adjacent tips on a tree (see Fig. \ref{Tree1}). Let $N_{\Lambda}(t|T)dt$ be the average number of cherry pairs on the tree of length $T$, separated by the time distance $t$. Then, if the first branch splits at time $T_1$ and the two resulting subtrees possess $M_1$ and $M_2$ leaves, respectively, one gets the distribution in the form
\begin{equation}
	N_{\Lambda}(t|T)=\left[2N_{\Lambda}(t|T-T_1)+
	 P^2(1|T-T_1) \delta\left(t-2\left(T-T_1\right)\right)I\left(0 \leq t\leq2T\right)\right]e^{-\mu T_1}.
	\label{CherriesMinimalDistanceEquation}
\end{equation} 
After average over $T_1$ and solving the resulting equation one obtains
\begin{equation}
	N_{\Lambda}(t|T)=\frac{ \lambda  (\lambda -\mu )^4 \sigma ^2}{2 }\frac{e^{\frac{ t \lambda }{2}+T \lambda +\frac{3 t \mu }{2}-T \mu } }{\left[e^{\frac{t \mu }{2}  } (\mu -\lambda  +\sigma  \lambda  )-e^{\frac{t \lambda }{2}  } \lambda  \sigma \right]^4}
	%I\left(0 \leq t\leq2T\right)
\end{equation}
for $0 \leq t\leq2T$ and 0 otherwise, resulting in Eq. \eqref{CherriesMinimalDistance}.

\section{The distribution of $N^1(t|T)dt$}
\label{AppNs}
In this Appendix we derive the distribution of $N^1(t|T)dt$. Consider an infinitesimal (in practice very small) interval, $[t,t+dt]$, such that $N(t|T)dt \ll 1$. The number of pairs $N^1(t|T)dt$ in this interval is distributed with the mean $N(t|T)dt$. The full distribution can be derived using the following arguments. 

Pairs, separated by the time in the interval $[t,t+dt]$, branched at the time interval $[T-t/2-dt/2,T-t/2]$. The probability to have a branch in this interval is given by $\lambda e^{(\lambda-\mu)(T-t/2)} dt/2$. Given that there is a branching point in this interval it can lead to different number of leaves and, therefore, pairs separated by the time in the interval $[t,t+dt]$. The probability that no observed pairs survive from this branching is given by
$
	1-[1-P(0|t/2)]^2,
$ 
where $P(n|T)$ is the probability to observe $n$ leaves on a tree of age $T$ and is given in Eqs. (\ref{P0},\ref{Pn}).
The probability that there are no observed pairs separated by the time in the interval $[t,t+dt]$ is given by Eq. \eqref{PrNs0}. 
The probability that there are $n>0$ observed pairs separated by the time in the interval $[t,t+dt]$ is given by
\begin{align}
\Pr\left(N^1(t|T) dt = n\right) &=\lambda e^{(\lambda-\mu)(T-t/2)} dt/2 
\sum_{n_1,n_2=1}^{n}  P(n_1|t/2)P(n_2|t/2) \delta_{n_1n_2,n}  \nonumber 
 \\  &= \lambda e^{(\lambda-\mu)(T-t/2)} dt/2 
\sum_{n_1 | n }  P(n_1|t/2)P(n/n_1|t/2).
\label{PrNs}
\end{align}
The last sum runs over all divisors of $n$, including $1$ and $n$.

\begin{figure}[tb]
\centering
\includegraphics[width= 0.5\columnwidth]{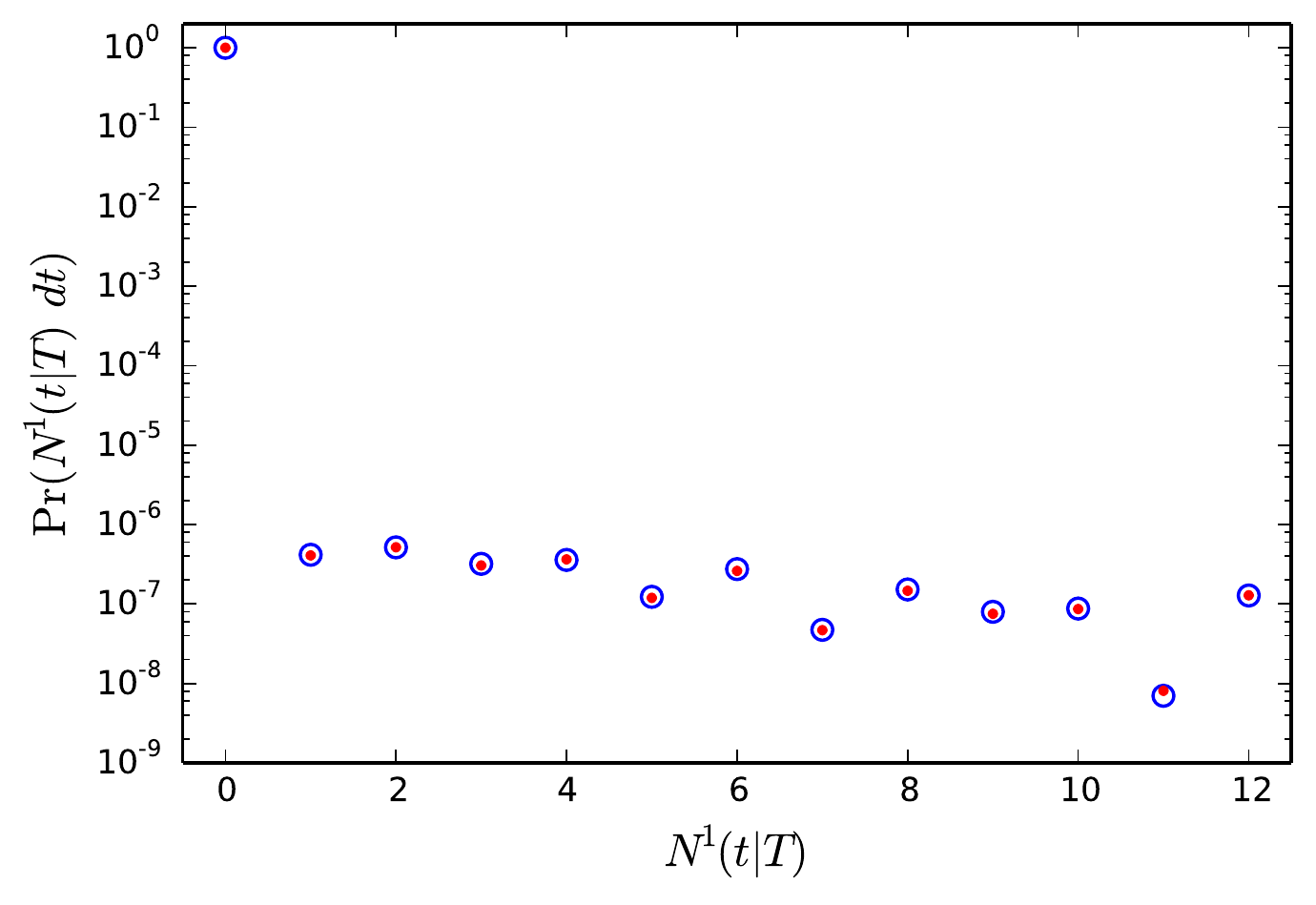}
\caption{Probability to observe a certain number of pairs separated by the time in the interval $[t,t+dt]$ on a tree of age $T$, $N^1(t|T) dt$. In this plot $T=1$, $\lambda=11$, $\mu=5$, $\sigma=0.01$, $t=1.5$ and $dt=0.00001$. Circles denote the results of numerical simulation and dots were obtained using the analytic formulas \eqref{PrNs0} for zero value and \eqref{PrNs} for non-zero values. Note the gap between zero and non-zero probabilities due to small bin size, $dt$.}
\label{NsDistFig}
\end{figure}

\bibliographystyle{apsrev}
\bibliography{Ref}%%%refs.bib

\end{document}